# A note on the notion of now and time flow in special relativity

Mario B. Valente

Pablo de Olavide University
Seville, Spain
mar.bacelar@gmail.com
Orcid: 0000-0001-9473-5005

Abstract: With special relativity, we seem to be facing a conundrum. It is a very well-tested theory; in this way, the Minkowski spacetime must be "capturing" essential features of space and time. However, its geometry seems to be incompatible with any sort of global notion of time. We might only have local notions of now (present moment) and time flow, at best. In this note, we will explore the possibility that a pretty much global notion of now (and time flow) might be hiding in plain sight in the geometry of the Minkowski spacetime.

The spacetime of special relativity – the Minkowski spacetime – seems to be incompatible, at least, with a universal notion of time, as is the case with the spacetime of Newtonian physics.

In fact, authors have even proposed that there is no real notion of now (present time), or some way to include in the geometry of spacetime what for us is the intuitive notion of the flow of time (see, e.g., Petkov 2009).

Others have tried to salvage at least a local notion of now and time flow (see, e.g., Dieks 1988, 2006; Arthur 2019; Savitt 2020).

A tentative work has tried to somewhat approach the universal inertial time of Newtonian spacetimes (Valente 2016a). We will take the final part of that work as the starting point of this brief note.

In that work, a global time is approached by considering an "extra hypothesis"; this corresponds to a setting of "initial conditions" in which the systems of interest are initially located at the "origin". This enables, e.g., to show that, in this case, the relativity of simultaneity is what some authors call a "kinematic effect", a sort of "make-belief" relativistic effect, like the case of the time dilation or the length contraction (see, e.g., Smith 1993). For example, in the case of time dilation, for an inertial "observer", the clocks of other inertial observers, in relative motion, go slower; it is like they go through less time when the observer, e.g., goes through one second. Now, all inertial observers measure the time passage of the other observers – in relative motion – as being smaller. It is not something that is really happening to any of them. The time dilation is due to the fact that the measurement of the time of a clock in relative motion by a clock taken to be at rest, is relative to this clock. It depends, e.g., on light being sent from this clock to the clock at "motion" and being sent back to the clock "at rest". This measurement procedure is made by taking a particular clock to be at rest and light being sent and received by this clock. It is a procedure that is "relative" to this clock; it is not a "universal" procedure independent of the adopted inertial observer (see, e.g., Bohm 1996).

Regarding the relativity of simultaneity, in Valente (2016a) it is addressed by considered the "extra hypothesis" that all physical systems in question coincide at the origin (see figure 1).

The usual interpretation of the relativity of simultaneity tells us that while for observer A the events in the worldlines of C and D, corresponding to $T_C$ and $T_D$, are simultaneous to A (occurring at time $T_A$), this is not the case for observer B. In B's case, the event in the worldline of D occurs at time $T_{B1}$ and the event on the worldline of C occurs at the *later time* $T_{B2}$. So, both observers A and B do not agree on what events corresponds to the present time. To say it a bit differently, each one has a different simultaneity plane (which is usually interpreted as the set of events that an observes takes to be simultaneous to him/her).

In the case of figure 1, we can see that this interpretation leads us astray.

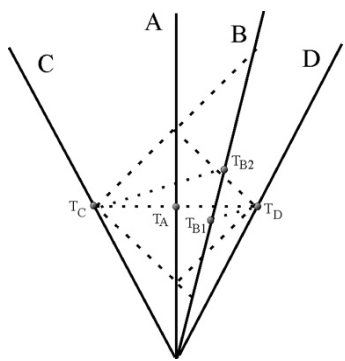

Figure 1. Another perspective on the relativity of simultaneity

As Valente writes:

> As it is well known, observer A will take the events along the worldlines of C and D corresponding, e.g., to $T_C = T_D = 10s$ to be simultaneous, having in A's coordinate system, e.g., the value $T_A = 12s$; due to the time dilation even if A takes the now-points of C and D to be simultaneous she attributes to them a value greater than their proper time readings. In the case of observer B he considers that the events along the worldlines of C and D corresponding to $T_C = T_D = 10s$ occur at the different moments $T_{B1}$ and $T_{B2}$. Regarding the relation between the temporal values of the now-points of C and D and A and B, both A and B are in a way wrong! When C and D go through 10 seconds, A and B also go through 10 seconds. (Valente 2016a, 31)

So, if we take $T_C$ and $T_D$ to be 10 seconds as measured by each of C and D's clocks, then they are simultaneous to A and B when each of them also measures 10 seconds; the values $T_A$, $T_{B1}$, and $T_{B2}$ would be an "artifice" of the measurement procedures that are "relative" to each inertial observer. The real time-relation between these four inertial observers is that they all measure the same inertial time – at the same time. So, if we take an event of say C's worldline corresponding to his/her clock measuring 10 seconds, then this event is *really simultaneous (or present)* to the events in A, B, and D's worldlines in which each of their clocks measures 10 seconds.[1]

While this work shows us that we can, by adopting the "initial conditions", approach in the Minkowski spacetime, a shared inertial time and some notion of shared now or present (and also of time flow), we are still far from the universal inertial time that we have in Newtonian spacetimes.

Adopting a four-dimensional formulation of Newton's space and time, we see that there is a natural "foliation" of this spacetime (see, e.g., Friedman 1983). We have a sort of stacking of three-dimensional spaces as time flows (see figure 2, for the case of one spatial dimension). All events of a three-dimensional space share the same inertial time.

However, in the case of special relativity, the existence of different simultaneity planes seems to make it impossible to have some sort of "stacking" of three-dimensional spaces as universal inertial time flows.

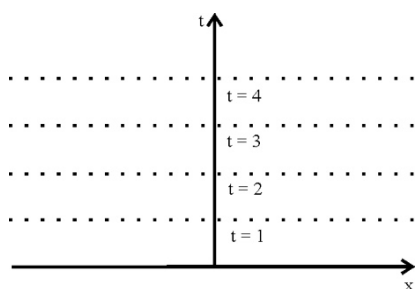

Figure 2. The foliation of Newtonian spacetime

---

[1] Another work where the relativity of simultaneity is addressed as a sort of kinematic effect is Valente (2013). In that work, the relativity of simultaneity is addressed in terms of the synchronization of clocks.

In the present note, we make the case that, contrary to the received view, there is, in fact, a "natural" stacking in Minkowski spacetime, which means that there is a shared universal inertial time, as is the case with Newtonian spacetimes. So, there is a global now and time flow in the Minkowski spacetime.

Let us say that we are considering a flat region of spacetime – the spacetime is locally Minkowskian. Now we have a myriad of inertial observers with different velocities sent in all directions from the same spatial location (in four-dimensional parlance, we have a spacetime event in which inertial observers are "emitted" with different velocities and directions). We will only consider one spatial dimension. The set of events corresponding to one "tick" of each observer's clock is a spatial (spacelike) "hyperbolic surface" represented by a hyperbola in a two-dimensional Euclidean diagram (see figure 3).[2] All the inertial observers have had the same passage of time: the flow of time is the same for all; also, all share the same now: the inertial time is the same for all of them.

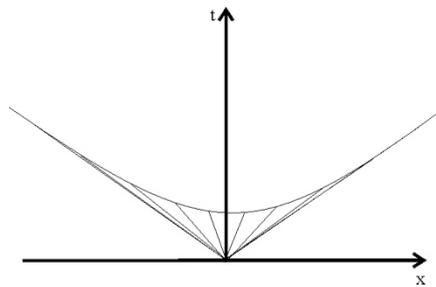

Figure 3. The hyperbola corresponding to one time unit of all inertial clocks

If we consider another "tick" of inertial time, we will have another non-Euclidean space stacked on top of the first, and so on (see figure 4). Contrary to the case of the length contraction, in which lengths are relative to an observer, a length along the arc of the hyperbola is not dependent on the adopted observer. That is, if instead of an inertial observer A we adopt an inertial observer B (moving relative to A) to be our observer "at rest", the relative distance along the hyperbola between, e.g., A and B is the same, as is the case of the distance between any other inertial observers (or more generally any events in the non-Euclidean space constituted by the hyperbolic surface).[3]

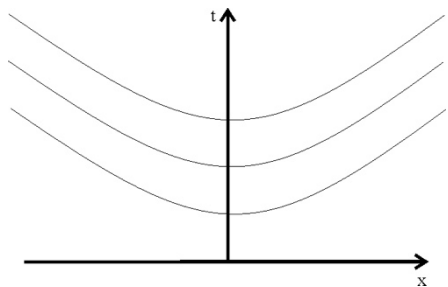

Figure 4. The foliation of the Minkowski spacetime

So, in each of the non-Euclidean spaces, there are universal distances (not relative to any adopted observer), and the time is the same for all observers (or, more generally, events on the surface). As the inertial time "unfolds" ("as time goes by"), we have a stacking of successive non-Euclidean spaces, exactly like in the case of Newtonian spacetime (the main difference is that in Newtonian spacetime the stacking is of Euclidean spaces and here of "hyperbolic" spaces).

---

[2] Regarding spacetime diagrams, it is important to take into account that we are representing in a Euclidean plane the geometry of the Minkowski spacetime. In this way, "a spacetime diagram is a projection of a two-dimensional section of spacetime with a geometry summarized by $\Delta s^2 = -(c\Delta t)^2 + \Delta x^2$ on the plane of a sheet of paper whose geometry is summarized by $\Delta s^2 = \Delta x^2 + \Delta y^2$" (Hartle 2003, p. 57).

[3] The distance between two events along a hyperbola (i.e., the Minkowski length of the arc of the hyperbola) is given by $\rho(\beta - \alpha)$, where $\alpha$ and $\beta$ are the hyperbolic angles of the events, and $\rho$ is the "radius" of the hyperbola (e.g., one, for the hyperbola corresponding to one unit of inertial time) (see, e.g., Dray 2012).

Before considering a criticism of this view, let us look at a simple example of this view "at work". The key is that we have to address the Minkowski diagrams in terms of the hyperbolas (see figure 5).

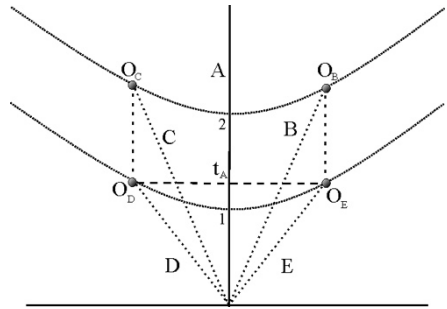

Figure 5. An example of the foliation of spacetime at work

In figure 5, we depict two hyperbolas corresponding to one "tick" of inertial time and two "ticks" of inertial time. The events $O_D$ and $O_E$, corresponding say to the "emission" of inertial clocks, are in the one "tick" hyperbola; they coincide with the events corresponding to observers D and E having their clocks ticking one unit of time. Their real time is not that attributed by A to them (time $t_A$); their real time is one, the same as that of observers D and E. In fact, the real time of A corresponding to these events is not $t_A$ (a time that is relative to A), but also one, as we can check in the hyperbola. Let us say that after a passage of time of one unit as measured by the emitted inertial clocks, they are "absorbed" by the inertial observers C and B (at spacetime events $O_C$ and $O_B$). What is the time of A, B, C, D, and E corresponding to these events? It is, checking with the second hyperbola, two units of time.

From the perspective of the relativity of simultaneity, events $O_D$ and $O_E$ are only simultaneous (present to each other) for observer A; the other observers in relative motion will consider that they occur at different times. However, these events are present with all the events of the one-hyperbola. In the same way, events $O_C$ and $O_B$ are present with all events of the two-hyperbola. All the inertial observers (or events) of each spatial non-Euclidean space (represented in the Euclidean diagram as a hyperbola) share the same present and all the inertial observers go through the same flow of time. For example, in the two-hyperbola, they all have gone through two units of time.

We will make the case that this is enough to consider that in the Minkowski spacetime there is a universal inertial time like in Newton's case; and, in this way, a clear global notion of now (or present) and of the flow of time.

To show this, let us address an argument by Dieks (2014), in which he criticizes a possible interpretation along the lines of what we are presenting here. According to Dieks, the events one time unit later than the event at the "origin" are not just the events in the hyperbola but all the events in the hyperbola and above it (see figure 6). In this way, Dieks considers that in the Minkowski spacetime, there is "no global time function defined on it" (Dieks 2014, 100).

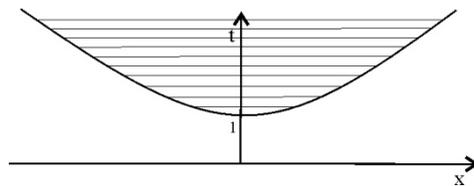

Figure 6. The events corresponding to the passage of one time unit according to Dieks (2014)

Dieks makes his argument by taking into account non-inertial motion. Let us consider, e.g., events O and A in figure 7, one unit of time "away" from each other, as measured by the inertial observer connecting these two events. If we now take an accelerated clock connecting these two events, the time ($t_1$) it measures is smaller than that of the inertial observer. If we consider another clock with a larger acceleration, the elapsed time ($t_2$) is even smaller (see figure 7).

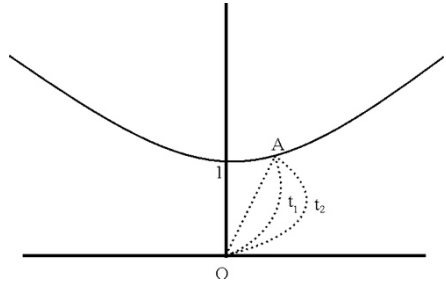

Figure 7. Timelike worldlines of an inertial and non-inertial clocks ($t_2 < t_1 < 1$)

This implies, as Dieks remarks, that "by traveling fast enough […] along a non-inertial path we can push the event at which one time unit has passed arbitrarily far into the future" (Dieks 2014, 101); i.e., further into the "shaded" area inside the hyperbola in figure 6. Dieks concludes that "the locus of events one time unit later than a given event does not define a sensible notion of simultaneity" (Dieks 2014, 101). This would imply that no meaningful foliation of the Minkowski spacetime is possible after all.

Let us counter-argue this with the help of figure 8. Again, the key is to take into account the successive hyperbolas (the one-hyperbola, the two-hyperbola, and so on).

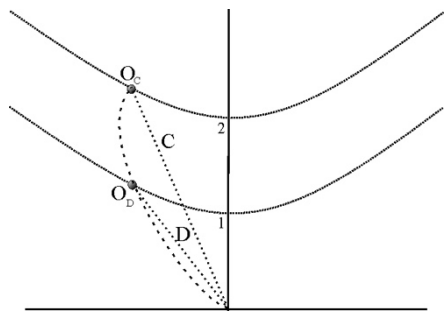

Figure 8. The worldline of a non-inertial clock superposed to the hyperbolas for one and two time units of inertial time (and the worldlines of two inertial clocks, C and D)

The elapsed time along a non-inertial path connecting two events is distinct from that of an inertial path connecting these same events. That is a given. However, every non-inertial timelike worldline is crossing a hyperbola. This means that independently of how time is elapsing for an accelerated clock, each event along the clock's worldline belongs to a "hyperbolic space" characterized by a universal inertial time. The "beating" of the accelerated clock is slower than that of the inertial clocks, but, e.g., in events $O_D$ and $O_C$ the "beating" is "happening" in a "hyperbolic space" corresponding first to one time unit and then to two time units. At event $O_D$ the accelerated clock (whatever its "beating" is) is simultaneous (present to/with) the clock of D that has just ticked one time unit; in the same way, in event $O_C$ the accelerated clock is present to/with the clock of C that has just ticked two time units. We can describe the non-inertial worldline in terms of the inertial time associated to each of the space-time events along the worldline (independently of how slowly the non-inertial clock is beating).[4]

---

[4] In fact, the accelerated clock' measured time – its proper time – is calculated in terms of the "underlying" inertial time. One way to see this is that an infinitesimal element of proper time is given by an infinitesimal inertial time interval multiplied by sqrt $(1 - v^2/c^2)$ (see, e.g., Valente 2016b, 2019). Another way is to consider light clocks (basically two mirrors, at a small distance, parallel to each other and exchanging light "pulses"). The propagation of light is closely related to inertial time. If two mirrors are a distance $d$ apart, taking into account the constant two-way speed of light $c$, as light is "emitted" by one mirror, bounces in the other, and returns to the first, "the inertial time associated with the propagation of light is $2d/c$" (Valente 2019, 17). Along a non-inertial worldline, we have "little" (infinitesimal) bounces of light corresponding to "little" (infinitesimal) intervals of inertial time (since we always have an inertial propagation of

In special relativity inertial observers are special; we should not put accelerated clocks at the same level as inertial clocks in our arguments.

Giving primacy to inertial clocks, arising from the primacy of inertial time (Valente 2019), we can see that there is a "natural" foliation of the Minkowski spacetime made in terms of a universal inertial time.[5]

---

light). We can see that "the total time read off by a light clock along a non-inertial worldline results from the (inertial) propagation of light; it is "built" from inertial time" (Valente 2017, footnote 19). The notions of inertial motion and time and of (inertial) propagation of light are more fundamental in the theory than the notion of clock (see Valente 2019).

[5] Notice that in this work we do not adopt any extra hypothesis like it is done in Valente (2016a). The adoption of inertial observers/clocks (and their coincidence at the origin) is simply an artifice that (hopefully) makes the presentation clearer to the readers. We can reframe the argumentation in terms of the geometry of the Minkowski spacetime and its physical interpretation. In particular, the existence of invariant non-Euclidean hyperbolic spaces for which the relative lengths between events are independent of the adopted inertial frame and for which there is a unique value of inertial time shared by all space-time events of a hyperbola (notice that a space-time event takes for us precedence over a particular physical system called clock). This enables a "natural" stacking/foliation of spacetime by hyperbolas. The present hyperbola (as the reader is reading this) corresponds to our space "now", at this present moment; as time flows, we have successive hyperbolas corresponding to successive "nows".